\documentclass[%
 aip,
 amsmath,amssymb,
 reprint,%
]{revtex4-1}

\usepackage{graphicx}
\usepackage{dcolumn}
\usepackage{bm}

\usepackage[utf8]{inputenc}
\usepackage[T1]{fontenc}
\usepackage{mathptmx}
\usepackage{etoolbox}
\usepackage{xcolor, soul}

\makeatletter
\def\@email#1#2{%
 \endgroup
 \patchcmd{\titleblock@produce}
  {\frontmatter@RRAPformat}
  {\frontmatter@RRAPformat{\produce@RRAP{*#1\href{mailto:#2}{#2}}}\frontmatter@RRAPformat}
  {}{}
}%
\makeatother
\begin{document}
\preprint{AIP/123-QED}

\title{Interface-sensitive microwave loss in superconducting tantalum films sputtered on c-plane sapphire}
\author{Anthony P. McFadden}
\email{anthony.mcfadden@nist.gov}
\affiliation{National Institute of Standards and Technology, Boulder CO, 80305}
\author{Jinsu Oh}
\affiliation{Ames National Labs, Ames, IA 50011}
\author{Lin Zhou}
\affiliation{Ames National Labs, Ames, IA 50011}
\affiliation{Department of Materials Science and Engineering, Iowa State University, Ames, IA 50011}
\author{Trevyn F.Q. Larson}
\author{Stephen Gill}
\author{Akash V. Dixit}
\author{Raymond Simmonds}
\author{Florent Lecocq}
\affiliation{National Institute of Standards and Technology, Boulder CO, 80305}

\date{\today}

\begin{abstract}
Quantum coherence in superconducting circuits has increased steadily over the last decades as a result of a growing understanding of the various loss mechanisms.
Recently, tantalum (Ta) emerged as a promising material to address microscopic sources of loss found on niobium (Nb) or aluminum (Al) surfaces.
However, the effects of film and interface microstructure on low-temperature microwave loss are still not well understood. Here we present a systematic study of the structural and electrical properties of Ta and Nb films sputtered on c-plane sapphire at varying growth temperatures. As growth temperature is increased, our results show that the onset of epitaxial growth of $\alpha$-phase Ta correlates with lower Ta surface roughness, higher critical temperature, and higher residual resistivity ratio, but surprisingly also correlates with a significant increase in loss at microwave frequency. Notably, this high level of loss is not observed in Nb films prepared in the same way and having very similar structure. By experimentally controlling the surface on which the Ta film is nucleated, we determine that the source of loss is only present in samples having an epitaxial Ta/sapphire interface and show that it is apparently mitigated by either growing a thin, epitaxial Nb inter-layer between the Ta film and the substrate or by intentionally treating, and effectively damaging, the sapphire surface with an \textit{in-situ} argon plasma before Ta growth. In addition to elucidating this interfacial microwave loss, this work provides adequate process details to aid reproducible growth of low-loss Ta films across fabrication facilities.
\end{abstract}

\maketitle

\section{Introduction}

In the field of superconducting quantum computing, the transmon -a Josephson junction (JJ) shunted by a capacitor\cite{Koch2007,Nakamura1999}- is a promising candidate moving towards useful quantum computation\cite{Krantz2019,Kjaergaard2020}. This technology has advanced rapidly, as qubit coherence times have benefited from improvements in measurement setup, device design layout\cite{Krantz2019,Kjaergaard2020}, and more recently advances in materials used to form the capacitor metal, where state-of-the-art transmon coherence times have been demonstrated using aluminum (Al)\cite{Biznarova2024}, niobium (Nb)\cite{Bal2024,Kono2024}, and tantalum (Ta)\cite{Place2021}. The capacitor surface and its interface with the substrate are known to host two-level systems (TLS) which act as dissipation channels which limit coherence times in superconducting qubits\cite{Simmonds2004,Siddiqi2021}.

Tantalum grown on widely available c-plane sapphire substrates has emerged as an especially promising candidate for low-loss superconductor-based capacitors. Ta was shown to offer coherence time improvements when compared to Nb in reports of state-of-the-art transmons made using Ta-metal capacitor pads shunted by Al/AlO$_x$/Al JJ's\cite{Place2021,Wang2022}. The reason for the enhanced performance in Ta devices has been proposed to be due to the more favorable properties of the surface oxides of Ta as compared to Nb\cite{Place2021,Crowley2023}. Optimizing the growth and processing conditions of the capacitor material for low microwave (MW) loss and understanding the mechanisms of surface and interface loss is presently an active area of research.

Superconducting devices made from Ta typically utilize $\alpha$-phase Ta formed in the body-centered cubic, (BCC) structure, which has a higher superconducting critical temperature ($T_c$) of 4.49K\cite{Inaba1980} as compared to the tetragonal $\beta$-phase Ta, which has a $T_c$ below 1K\cite{Mazin2022}. For Ta thin films, room temperature deposition usually results in the metastable $\beta$-phase, while formation of $\alpha$-phase Ta typically requires heating the substrate during growth\cite{Jiang2005} or deposition on a seed layer\cite{Lee1998,Face1987,Urade2024,Alegria2023}. This formation of two distinct elemental phases contrasts with Nb, for which metastable phases do exist, but thin films are most commonly found in the BCC structure\cite{Lee2023,Kunzmann2024}. Notably, the lattice parameters of Ta and Nb in the BCC structure are nearly the same, having values of 0.3306 nm and 0.3301 nm respectively\cite{Ward2003}.

The goal of reducing MW loss originating in the transmon capacitor metal and its interfaces is to increase qubit coherence times; however, disentangling the capacitor metal contribution to overall decoherence rate is nontrivial. Transmon fabrication requires steps to form the JJ, which may introduce additional loss channels, both due to the junction itself as well as residue, contamination, or other unwanted effects of processing. When optimizing transmon design layout for high coherence times, the capacitors are intentionally made large (mm-scale) in order to dilute the electric field participation at surfaces and interfaces which are thought to be the dominant contributors to decoherence\cite{Wang2015,Gambetta2017}.

Increasing the size of the capacitor also increases the mode volume in the packaging environment, where care must be taken to avoid package related loss from lossy dielectrics and resistive (non-superconducting) metals in the package. In addition, resonant modes present in the MW environment of the package can also couple to the qubit and act as decoherence sources, limiting coherence times to below state-of-the-art\cite{Huang2021,Sheldon2017}. Measurement of high coherence transmons also requires substantial shielding and filtering in the measurement setup to suppress the quasiparticle-generating radiation seen by the qubit that also limits coherence times\cite{Gordon2022}.

In contrast, coplanar waveguide (CPW) resonators may also be used to assess MW loss arising from the superconducting metal layer and its interfaces while avoiding extra processing steps, measurement difficulties associated with transmon characterization, and additional loss associated with Josephson elements\cite{McRae2020b}. Resonator internal quality factor ($Q_i$), a figure of merit expressing the dissipation rate due to materials and environmental-related losses in the resonator, may be measured using a continuous wave measurement by fitting the reflected ($S_{11}$) or transmitted ($S_{21}$) resonance spectra to established models\cite{McRae2020b}. Reducing the CPW resonator gap width confines the mode volume and increases the electric field participation at the metal surfaces and interfaces. Because MW loss is typically dominated by surface and interface contributions, reducing CPW gap width tends to reduce $Q_i$, while also reducing package-related contributions to MW loss and providing an exacting evaluation of the superconducting metal, surface, and interface.

In this work, the structural and electrical properties of Nb and Ta films grown on c-plane sapphire by sputtering have been studied as a function of growth temperature while including sufficient details to provide a starting point for reproducibility. The MW properties of the films were evaluated using narrow-gap CPW resonators, transport properties were measured using temperature dependent resistivity measurements, and the structure of the films was characterized using X-ray diffraction (XRD), atomic force microscopy (AFM), reflection high energy electron diffraction (RHEED), and transmission electron microscopy (TEM). 

Measurements of these films revealed an unknown source of MW loss in epitaxial Ta(111) films grown directly on sapphire. An experiment was devised to test the Ta/sapphire interface by forming inter-layers of either epitaxial Nb(111) or amorphized sapphire between the crystalline substrate and the Ta films, which suggests that the epitaxial $\alpha$- Ta(111)/ Al$_2$O$_3$(0001) interface is a significant source of MW loss and that loss may be avoided by simple treatments to the substrate surface before growth. While the suggested inter-layers have been employed in high performing superconductive circuits \cite{Urade2024,Alegria2023}, a systematic evaluation of their effects on structural and electrical properties and a clear comparison to epitaxial $\alpha$-Ta(111)/Al$_2$O$_3$(0001) films has not yet been shown in the literature.

\section{Methods}

\subsection{Materials Growth and Micro-fabrication}
As-received, 76.2mm diameter, single-side-polished Al$_2$O$_3$ (0001) wafers were prepared by solvent cleaning in acetone and isopropanol with ultrasonication followed by a spin-rinse-dry (deionized water rinse with $N_2$ dry) cycle and immediately loaded into the high-vacuum load lock of a sputter deposition tool. 

Wafers were secured to a silicon backing wafer and inconel sample holder using clips made from Ta wire. The silicon backing wafer was heated radiatively by a standard bank of quartz halogen lamps through a window in the inconel holder which was found to result in reproducible and uniform heating across the wafer surface. Variation between substrate surface temperature and measured thermocouple temperature can be significant, and is known to vary depending on details including the substrate holder and sample mounting technique\cite{Mizutani1988,Lee1991}. For the sake of reproducibility across different labs, tools, and sample mounting configurations, the thermocouple temperature was calibrated to the sample surface temperature at the melting point of aluminum (T=660$^\circ C$). This was done by depositing 500nm of aluminum on a bare sapphire wafer at room temperature and observing the phase transition by RHEED and visually through a viewport as the sample was slowly heated. The melting point of Al on the sapphire surface was found to occur at a thermocouple temperature of 790$^\circ C$ establishing that the surface temperature is approximately 130$^\circ C$ lower than the thermocouple reading around the Al melting point. Through the remainder of this work, "growth temperature" refers to the temperature read on the thermocouple, but it is important to note that the sample surface temperature is significantly lower.

The mounted sapphire substrates were annealed in the high-vacuum growth chamber for 2 hours at 880$^\circ C$ thermocouple temperature before cooling to the film growth temperature without breaking vacuum. This high vacuum annealing step was performed to reduce the amount of water and adventitious carbon on the sapphire surface\cite{Wilson2017}, providing a consistent substrate surface chemistry for all samples in this study.

Ta and Nb films were deposited at varying temperatures between room temperature and 730$^\circ C$ thermocouple temperature using DC magnetron sputtering at a pressure of 3 mTorr with 15 sccm Ar flow from 2-inch-diameter Ta or Nb targets having a metals basis purity of 99.999\%. A 300 W sputtering power was used, resulting in deposition rates near 1.7 \AA /s for both Ta and Nb, which was monitored using a quartz crystal microbalance calibrated to film thickness measured using X-ray reflectivity. After growth, samples were cooled to room temperature overnight in the growth chamber and removed from vacuum.

Following Ta or Nb deposition, the metal films were patterned into coplanar waveguide (CPW) resonators and Hall bars using standard photolithography and dry etching. The Ta or Nb coated wafer was first solvent cleaned to obtain a consistent starting surface, followed by spin coating of a heximethyldisilazane (HMDS)-based primer and standard photoresist. The wafers were patterned using direct laser writing, descummed in an oxygen plasma, and etched in an inductively coupled plasma (ICP) tool using the same chlorine/boron trichloride (Cl$_2$/BCl$_3$) etch chemistry for both Nb and Ta\cite{Bal2024}. After etching, the photoresist was removed using an n-methylpyrrolidone (NMP)-based remover held at 80$^\circ C$ with ultrasonication followed by an isopropanol rinse. The finished wafer was again coated in photoresist to protect the surface during dicing. 7.5mm x 7.5mm diced chips were cleaned in NMP-based remover at 80$^\circ C$ with ultrasonication followed by an isopropanol rinse. CPW resonator chips were additionally cleaned in 6:1 buffered hydrofluoric acid (BHF) for 2 minutes followed by a deionized water rinse immediately before packaging and wiring.

Notably, the substrate preparation procedure used for this study is minimal, avoiding the acid or plasma etching steps and furnace annealing commonly used with sapphire substrates \cite{Dwikusuma2002,Vlasov2016,Place2021}. Initially, samples were prepared by acid cleaning for 10 minutes in a stabilized formulation of sulfuric acid and hydrogen peroxide compounds held at 80$^\circ C$ (Piranha etch). This step was not found to have an appreciable effect on the resulting film structure or electrical properties, presumably because the as-received wafers were already clean. However, chips that were prepared from the same as-received wafers by covering with protective photoresist before dicing were found to have surface contamination after the resist was stripped which affected film growth. For these chips, a Piranha-type etch was found to result in a clean starting growth surface. Piranha etching is a recommended step in order to reproduce the results presented here. The samples studied in the main text of this work did not undergo a Piranha clean, though some additional results using Piranha cleaning are included in Appendix A.

\subsection{Structural Characterization}
All samples presented in this work were characterized using \textit{in-situ} RHEED, and \textit{ex-situ} AFM and XRD. These techniques are complementary: RHEED is sensitive to the first few monolayers of the surface and was used to examine the structure of the sapphire surfaces before growth and the metal surfaces immediately following growth. \textit{Ex-situ} AFM was used to examine the surface morphology of the metal films and XRD was used to study their crystalline phases and orientations as well as their epitaxial relationships to the substrate. 

For the results presented in this work, a 30kV electron source is used for RHEED, AFM was performed in tapping mode, and the X-ray diffractometer (Cu $K\alpha1$) was configured with a monochromator (2-bounce Ge(220)) installed on the incident beam only. Prior to all XRD measurements, angle offsets resulting from imperfect sample mounting in the diffractometer were established by aligning the instrument to the (0006) reflection of the Al$_2$O$_3$(0001) substrate.

Cross-sectional scanning transmission electron microscopy (STEM) investigation was performed to examine the microstructure of the films. The STEM samples were prepared using a focused ion beam instrument and then investigated with an aberration-corrected STEM at 200kV. A high-angle annular dark-field (HAADF) detector was used for dark-field imaging in STEM mode, with a convergent semi-angle of 18mrad and a collection semi-angle of 74–200mrad. Annular bright-field (ABF) images were acquired with a collection semi-angle of 11–19mrad. Energy-dispersive X-ray spectroscopy (EDS) was carried out with a probe current of 150 pA using an EDS detector.

\subsection{Electrical Measurements}
Characterization of Hall bars was performed in a helium-4 flow cryostat. A standard lock-in configuration was used to obtain film sheet resistance as a function of temperature using the voltage output of the lock-in amplifier operated at 87 Hz in series with a $10k\Omega-100k\Omega$ bias resistor. The series combination of the device under test and the cryostat wiring is under $10\Omega$, allowing the approximation of the voltage source/bias resistor combination as a current source. Excitation currents of 1-100 $\mu$A were used in a 4-point configuration. Residual resistivity ratio (RRR), a commonly used figure of merit for metallic thin films, is defined as the resistance at 300K divided by the resistance just above the superconducting transition temperature ($T_c$).

Microwave loss of deposited films was studied using CPW resonators. Each CPW resonator chip consists of 8 inductively coupled, quarter-wave resonators in hanger configuration off a central feedline similar to the design presented by Kopas et al. \cite{kopas2022}. The CPW dimensions of conductor/gap are 6$\mu$m/3$\mu$m and resonant frequencies are designed to fall between 5.5-6.5 GHz. The narrow conductor and gap widths are chosen to increase the electric field participation at surfaces and interfaces to study their contribution to microwave loss \cite{Gao2008,McRae2020b}. The electric field energy participation ratio of the metal/substrate interface for this CPW geometry is estimated to be between 0.4-0.5\% based on the works of others \cite{Calusine2018,kopas2022} where a 2nm thick interfacial layer was assumed. The electric field participation at the surfaces and interfaces of CPW resonators is more thoroughly discussed in other works \cite{McRae2020b,kopas2022,Calusine2018,Crowley2023}. Notably, in order to compute such a participation ratio, assumptions must be made about the extent and dielectric properties of interfacial regions which may not be well understood.

Clean CPW resonator chips were wire-bonded with aluminum wire and packaged in a 2-port package constructed of gold-plated copper for microwave measurements in a dilution refrigerator with a base temperature of 35mK. The attenuation and amplification setup used is similar to that presented by McRae et al.\cite{McRae2020b}. Microwave loss of the resonators was characterized using a vector network analyzer (VNA) by fitting the power dependent $S_{21}$ spectra of the resonators to the diameter correction method (DCM)\cite{Khalil2012,McRae2020b} model to extract internal ($Q_i$) and external ($Q_c$) resonator quality factors. A Josephson parametric amplifier\cite{Aumentado2020} installed at the 35mK stage was used for all low-power measurements (below 1000 photon occupation in the cavities).

Internal quality factor of resonators is commonly measured as a function of drive power and temperature in order to extract two-level system (TLS) loss and characterize materials\cite{McRae2020b, Crowley2023}.For the worst performing films in this work (epitaxial Ta grown directly on Al$_2$O$_3$(0001)), resonator $Q_i$ is markedly poor and changes by less than 25\% with power in contrast with the other samples presented where $Q_i$ can vary by as much as two orders of magnitude over a similar input power range. For this reason, internal quality factor at low power only is used as a performance metric to compare films. This avoids consideration of these saturable sources of loss which are generally of interest to the community, but are beyond the scope of this work, while focusing on the low power metric that is most relevant to transmon performance. More discussion of quality factor power dependence is included in Appendix E.

\section{Results and Discussion}

\subsection{Overview of Experiment}

Firstly, structural and electrical characterization results are shown for Ta and Nb films sputtered directly on clean, crystalline Al$_2$O$_3$(0001) surfaces at varying substrate temperatures. These foundational results establish that: 
\begin{enumerate}
    \item Structurally, both Nb and Ta films grown at varying temperature in this work reproduce prior results from detailed growth studies\cite{Wildes2001,Read1965,Feinstein1973} which is elaborated in Section \ref{TempdepSection}.
    \item Temperature dependent resistivity measurements of the Ta films grown at high temperature suggest high quality films. $T_c$ of epitaxial Ta films were measured to be near reported bulk values with RRR > 30 for 100nm thick films which is included in Table \ref{filmData_Ta}.
    \item The same microfabrication process is used for all samples presented here and is shown to yield high quality CPW resonators for all Nb films, but only for Ta films grown at lower temperature as shown in Figure \ref{NbandTaQi}. Ta films grown at higher temperature are epitaxial and high-quality structurally, but yield lossy resonators, seemingly at odds with structural and DC electrical characterization results.
\end{enumerate}

In light of these observations, subsequent samples were fabricated to test the Ta/sapphire interface. The details of this experiment are provided in Section \ref{modInterfaceSec}, and the results suggest that the epitaxial Ta/sapphire interface is the dominant source of microwave loss at both high and low power in the samples having low $Q_i$. While our experiments identify the interface as the source of loss and provide techniques shown to eliminate it, the microscopic mechanism of the loss is left as an open question.

\subsection{\label{TempdepSection}Temperature dependent growth of Nb and Ta on $\bf{Al_2 O_3 (0001)}$}

\begin{figure*}[hbt!]
\includegraphics[width=1.0\textwidth]{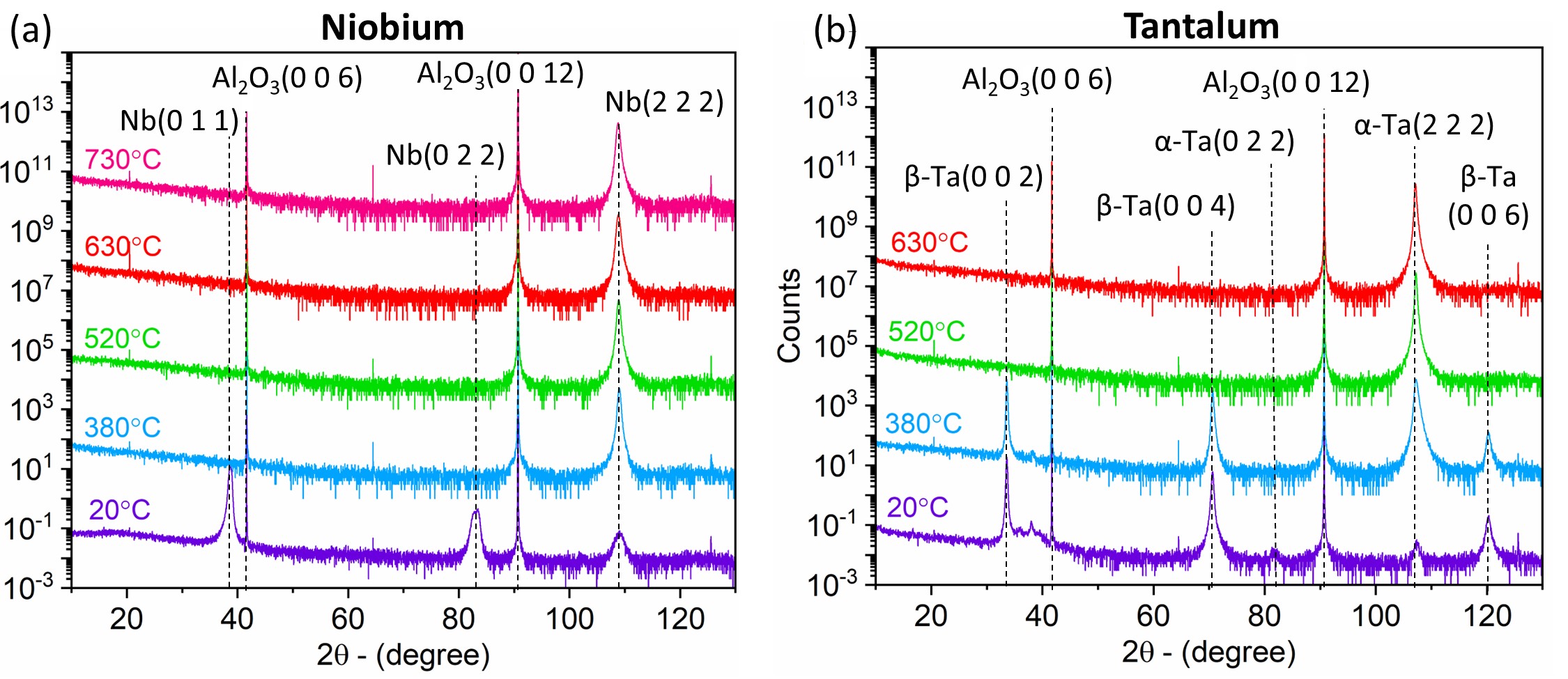}
\caption{\label{NbTa_XRD} Coupled ($\omega-2\theta$), on-axis (specular), X-ray diffraction measurements of 100nm thick films of (a) Nb and (b) Ta grown on $Al_2 O_3 (0001)$ at varying growth temperatures. Data are offset for clarity.}
\end{figure*}

Figure \ref{NbTa_XRD} shows XRD measurement results taken from Nb and Ta films grown at varying substrate temperature. Each peak indicates a crystalline phase oriented perpendicular to the substrate surface, along the growth direction. For Nb, growth at room temperature results in a polycrystalline film having both (110) and (111) oriented grains along the growth direction. Non-specular measurements were also performed which show that the (111) and (110) grains have in-plane texture. XRD pole figures and more discussion are included in Appendix C. For Nb films grown at elevated temperature, only epitaxial Nb(111) is observed with no notable changes to structure as the growth temperature is increased in the temperature range studied. Throughout this work, epitaxial implies that the grown film has a single crystallographic orientation both out-of-plane and in-plane which is aligned to the underlying substrate. Non-specular measurements on these films confirm the expected epitaxial relationship\cite{Mayer1992} $Nb(111)||Al_2 O_3 (0001)$ with $Nb[1 1 \bar2]||Al_2 O_3 [1 0 \bar1 0]$.

XRD results from the Ta films (Fig. \ref{NbTa_XRD}(b)) are somewhat more complex owing to the tendency of Ta thin films to crystallize in both the BCC ($\alpha$-Ta) and tetragonal ($\beta$-Ta) structures. Ta grown at room temperature results in a film consisting mostly of highly textured $\beta$-Ta(001) with a detectable amount of $\alpha$-Ta. In addition to out-of-plane texture, the $\beta$-Ta(001) was found to be textured in-plane. Non-specular measurements (not included) show three rotational domains of 4-fold symmetric $\beta$-Ta(001) aligned in-plane to the substrate, consistent with the 12-fold symmetric RHEED pattern (not included) observed immediately following growth. Ta deposition at 380$^\circ C$ resulted in a biphasic film having both textured $\beta$-Ta(001) and epitaxial $\alpha$-Ta(111) components, while growth at temperatures above 500$^\circ C$ resulted in only epitaxial $\alpha$-Ta(111) having the same structural relationship to the substrate as the epitaxial Nb(111) described previously.

Sheet resistance at 300K, RRR, and $T_c$ were extracted from temperature dependent resistance measurements for the Nb and Ta films and is included in Tables \ref{filmData_Nb} and \ref{filmData_Ta} respectively. For Nb films, RRR increases with growth temperature as the film structure changes from textured Nb(110) to epi-Nb(111), but is somewhat reduced for the highest growth temperatures.

\begin{table}[h]
\begin{ruledtabular}
\begin{tabular}{r|ccccc}

Nb Growth Temp. ($^\circ C$) & 20 & 380 & 520 & 630 & 730\\ \hline
RMS Roughness ($Angstrom$) & 9.1 & 2.8 & 3.5 & 6.5 & 8.1\\
$R_{sheet}$ at 300K ($\Omega / \square $)& 2.44 & 1.96 & 1.92 & 1.94 & 1.95\\
RRR& 5.1 & 45 & 75 & 65 & 66\\
$T_c$ ($K$)& 9.02 & 9.47 & 9.42 & 9.42 & 9.38\\
Resonator mean $Q_i$ (Low-power)& 340k & 270k & 520k & 440k & 140k\\

\end{tabular}
\end{ruledtabular}
\caption{\label{filmData_Nb}Measured properties of 100nm thick Nb films grown at varying temperature. The RMS roughness was obtained from a 500nm x 500nm scan}
\end{table}

For the textured $\beta$-Ta(001) film grown at 20$^\circ C$, resistivity \textit{increased} with decreasing measurement temperature (RRR<1) and $T_c$ was below the base temperature of the He-4 cryostat used in the measurement. Temperature dependent transport of the epitaxial $\alpha$-Ta(111) samples was consistent with high quality films. A RRR value of 29 was measured for the $\alpha$-Ta(111) grown at 520$^\circ C$ which increased to 47 for the film grown at 630$^\circ C$. The $T_c$ of the $\alpha$-Ta(111) films range from 4.24K-4.35K, near reported values of pure, bulk Ta (4.49K)\cite{Inaba1980}. For the epitaxial Ta films, RRR is improved with increasing thickness. Appendix A includes data from films not included in the main text, with the best 200nm thick epitaxial Ta film reaching RRR=78.

AFM measurements were also performed on all of the films studied. While changes in morphology were observed with changing growth temperature, all films were found to have root-mean-square (RMS) roughness values below 2nm and no strong correlation to transport properties was observed. RMS roughness values for both Nb and Ta are included in Tables \ref{filmData_Nb} and \ref{filmData_Ta} respectively. Additional atomic force micrographs and discussion is included in Appendix D.

\begin{table*}[ht!]
\begin{ruledtabular}
\begin{tabular}{r|cccccc}

Growth Temp. ($^\circ C$) & 20 & 380 & 520 & 630 & 630 & 630\\
Substrate Surface& clean $Al_2O_3$ & clean $Al_2O_3$ & clean $Al_2O_3$ & clean $Al_2O_3$ & 5nm Nb(111) / clean $Al_2O_3$  & Ar-treated $Al_2O_3$ \\
Phase & mostly $\beta$-Ta & mixed $\beta$ and $\alpha$-Ta & $\alpha$-Ta & $\alpha$-Ta(Sample A) & $\alpha$-Ta (Sample B)& $\alpha$-Ta (Sample C)\\ \hline

RMS Roughness ($Angstrom$) & 3.2 & 15.1 & 5.0 & 3.5 & 2.1 & 13.3\\
$R_{sheet}$ at 300K ($\Omega / \square $)& 77.3 & 2.53 & 1.51 & 1.31 & 1.39 & 1.45\\
RRR& $<$1 & 4.0 & 29 & 47 & 45 & 9.9\\
$T_c$ ($K$)& $<$1.8 & 4.01 & 4.24 & 4.35 & 4.48 & 4.21\\
Resonator mean $Q_i$ (Low-power)& 400k & 500k & 7.4k & 8.3k & 520k & 370k\\

\end{tabular}
\end{ruledtabular}
\caption{\label{filmData_Ta}Measured properties of 100nm Ta films grown at varying temperature.  The RMS roughness was obtained from a 500nm x 500nm scan}
\end{table*}

\begin{figure}[hbt!]
\includegraphics[width=0.5\textwidth]{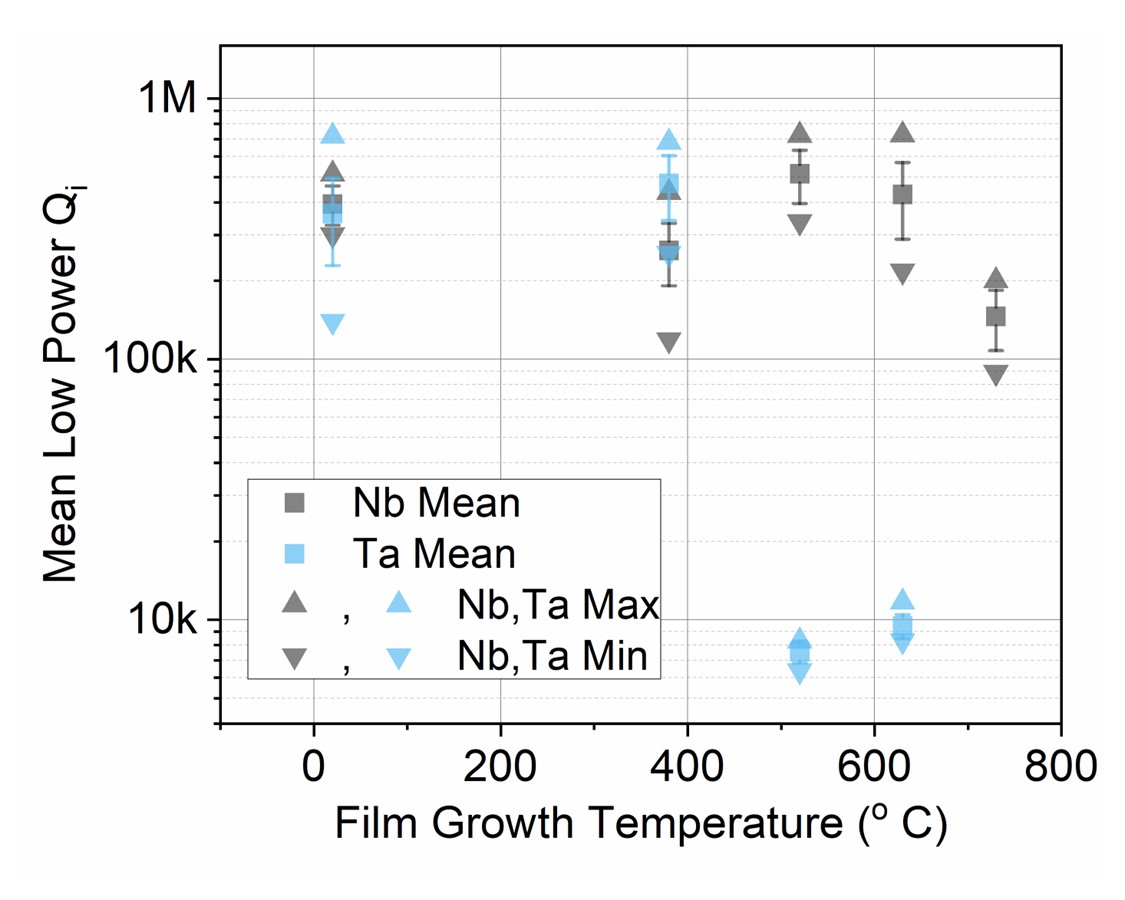} 
\caption{\label{NbandTaQi} Low power, average internal quality factor ($Q_i$) of CPW resonators fabricated from Nb and Ta films plotted against growth temperature. Mean, minimum, and maximum for each sample are indicated with the error bar representing the standard deviation. Each data point includes measurements taken from 3-8 different resonators.}
\end{figure}

Figure \ref{NbandTaQi} shows $Q_i$ of CPW resonators plotted against growth temperature for both Ta and Nb films. To obtain the data shown, $S_{21}$ spectra were measured for at least 3 resonators on each chip at varying input power corresponding to $10^6$ to below $0.05$ photons in the CPW cavities. $Q_i$ values taken below 0.5 photon occupation (at least 3 per resonator) were then averaged together and plotted in Fig. \ref{NbandTaQi}. Thus, each data point shown includes multiple measurements of low-power $Q_i$ from 3-8 different resonators measured from the same wafer, with the minimum, maximum, mean, and standard deviation indicated. These low-power $Q_i$ data may be considered as figures of merit that are representative of the material and reflect the degree of MW loss at the ultra-low electric field intensity where transmons operate. The number of resonators measured on each chip, number of low power measurements on each resonator, and the range of fit parameters obtained from these measurements may be found in the Supplementary Information for all samples included in the main text.

Measurements of the Nb resonators show that a maximum $Q_i$ value near 500k is obtained for the epitaxial films grown at intermediate growth temperatures near 500$^\circ C$. For growth at 730$^\circ C$, which was the highest growth temperature tested, resonator $Q_i$ is degraded. This reduction in $Q_i$ may be due to increased impurity incorporation in the Nb during growth due to chamber outgassing or intermixing at the Nb-sapphire interface.

Measurement results from the Ta resonators are more interesting and somewhat unexpected. Firstly, it was found that resonators from the $\beta$-Ta films grown at room temperature are high performing, with $Q_i$ values exceeding those made from Nb resonators grown using the same conditions. Resonators containing $\beta$-Ta were also shifted to lower frequency, presumably due to the kinetic inductance of the higher resistivity $\beta$-Ta film \cite{Zmuidzinas2012}.

Most importantly, and in contrast to Nb, the microwave performance of the epitaxial $\alpha$-Ta(111) was significantly reduced, with $Q_i$ of resonators made from these films nearly 2 orders of magnitude lower than Ta grown at lower temperature and all Nb films presented in this study. For these lossy, epitaxial Ta(111) films, $Q_i$ was essentially power-independent, suggesting that the source of loss is not saturable TLS\cite{McRae2020b}. Because the Ta grown at lower temperature showed significantly higher $Q_i$ than films grown at high temperature, one could consider the Ta(111) crystallographic orientation, film-substrate reactions, and/or the electronic structure of the interface to be potential sources of the MW loss.

While the markedly lower $Q_i$ of the two epi-Ta films is apparent in Fig. \ref{NbandTaQi}, variations between the other samples presented is less clear. Notably, several films in the data presented have maximum $Q_i$ values falling in the range 600k-800k, suggesting that the resonator performance may be limited by something common to them all, which includes the sapphire surface exposed to the dry etch when the Nb or Ta was etched away, the sapphire itself, and the backside of the substrate. A detailed analysis of the power dependent quality factor data with more measurements to provide the needed statistics may clarify optimum growth temperature for these films, though that is beyond the scope of this work.

\subsection{\label{modInterfaceSec}Modifying the Ta/$\bf{Al_2 O_3 (0001)}$ interface}

To further investigate the source of MW loss in the epi-$\alpha$-Ta(111) samples, the Ta/Al$_2$O$_3$(0001) interface was tested in two different experiments. Each experiment only differs in one aspect from the reference Ta film grown at 630$^\circ C$, referred to as Sample A in the following. 

For the first test, referred to as Sample B, a 5nm thick Nb interlayer was deposited on the sapphire prior to a 95nm thick Ta layer. Both layers were deposited on Al$_2$O$_3$(0001) prepared in the same manner described previously at a growth temperature of 630$^\circ C$, resulting in epitaxial growth of both the Nb and Ta which have the same crystal structure and nearly the same lattice constant. This sample provides an effective test of the epi-$\alpha$-Ta(111) interface because the Nb(111)/Al$_2$O$_3$(0001) interface was already shown to yield low-loss resonators and the Ta film grows in the same epi-$\alpha$-Ta(111) structure that was shown to yield lossy resonators when grown directly on the substrate. 

For the second test, referred to as Sample C, the Al$_2$O$_3$ (0001) substrate was treated with an argon (Ar) plasma in the growth chamber \textit{in-situ} at room temperature before the high vacuum outgas and growth of Ta directly on the substrate at 630$^\circ C$. The plasma exposure was performed at 10mTorr using an RF power of 50W for 120 seconds. This plasma condition was found to etch aluminum at 1.6 nm/min in a separate calibration that is provided here for repeatability. The outgas temperature (880$^\circ C$) used is not expected to result in restructuring of the Al$_2$O$_3$(0001) surface, which occurs at temperatures above 1000$^\circ C$ \cite{Wildes2001}.This plasma treatment induces damage to the sapphire surface, and the resulting Ta film was found to be polycrystalline and textured in XRD measurements. The damage to the sapphire surface was observed with RHEED and more details are included in Appendix B.

Figure \ref{TaonlyQi} shows low-power $Q_i$ measurements of 3 resonators measured from each chip selected from Samples A, B, and C. Growth of an epi-Nb inter-layer (Sample B) and Ar plasma treatment (Sample C) both result in a marked enhancement of resonator $Q_i$ by nearly two orders of magnitude, confirming that the epi-$\alpha$-Ta(111)/Al$_2$O$_3$(0001) interface is the source of loss in the worst performing samples. 

\begin{figure}[hbt!]
\includegraphics[width=0.48\textwidth]{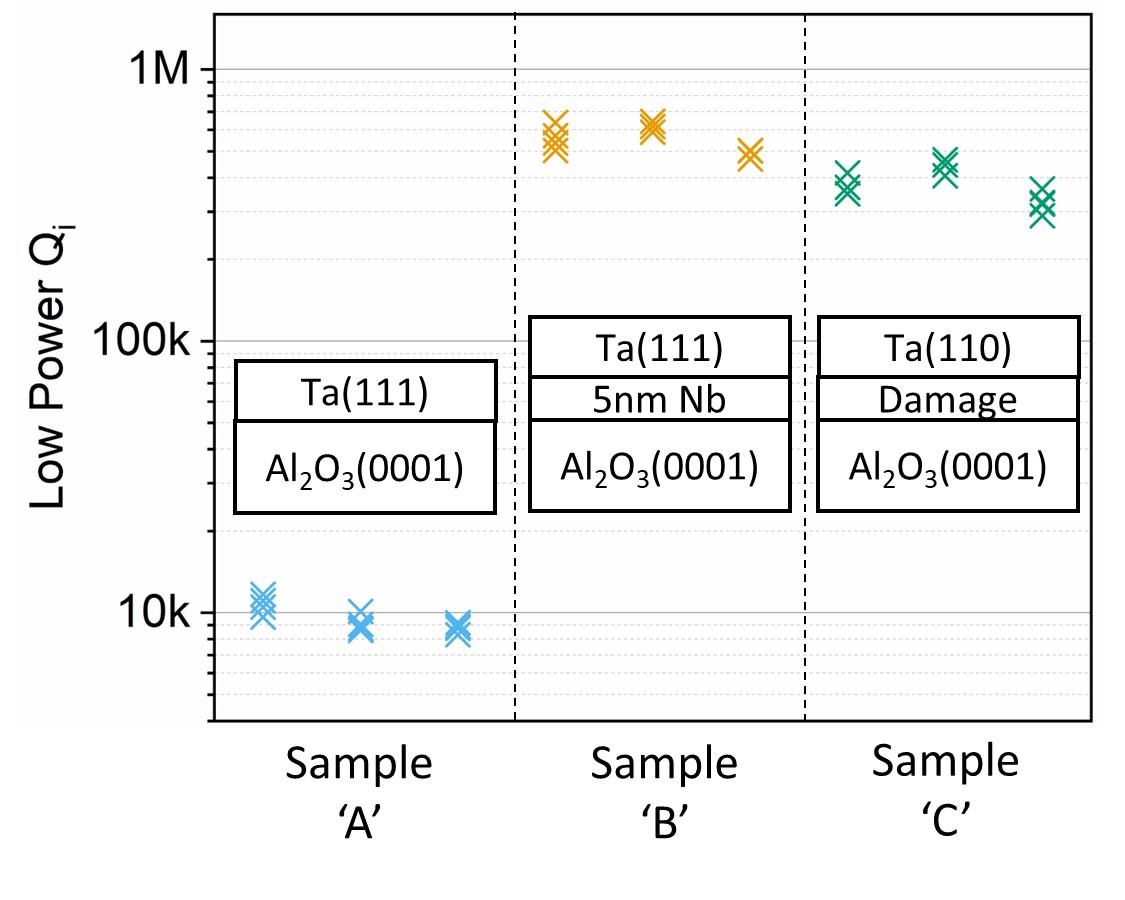} 
\caption{\label{TaonlyQi} Low power internal quality factor ($Q_i$) of CPW resonators fabricated from Samples A, B, and C. Each mark denotes a single measurement taken below 0.5 photon occupation in the resonator with data from 3 different resonators on each sample shown. Ta growth on both the epitaxial $Nb(111)$ interlayer (Sample B) and plasma-treated sapphire (Sample C) show significantly enhanced $Q_i$ as compared to the epitaxial films grown directly on sapphire (Sample A).}
\end{figure}

\begin{figure*}[hbt!]
\includegraphics[width=1.0\textwidth]{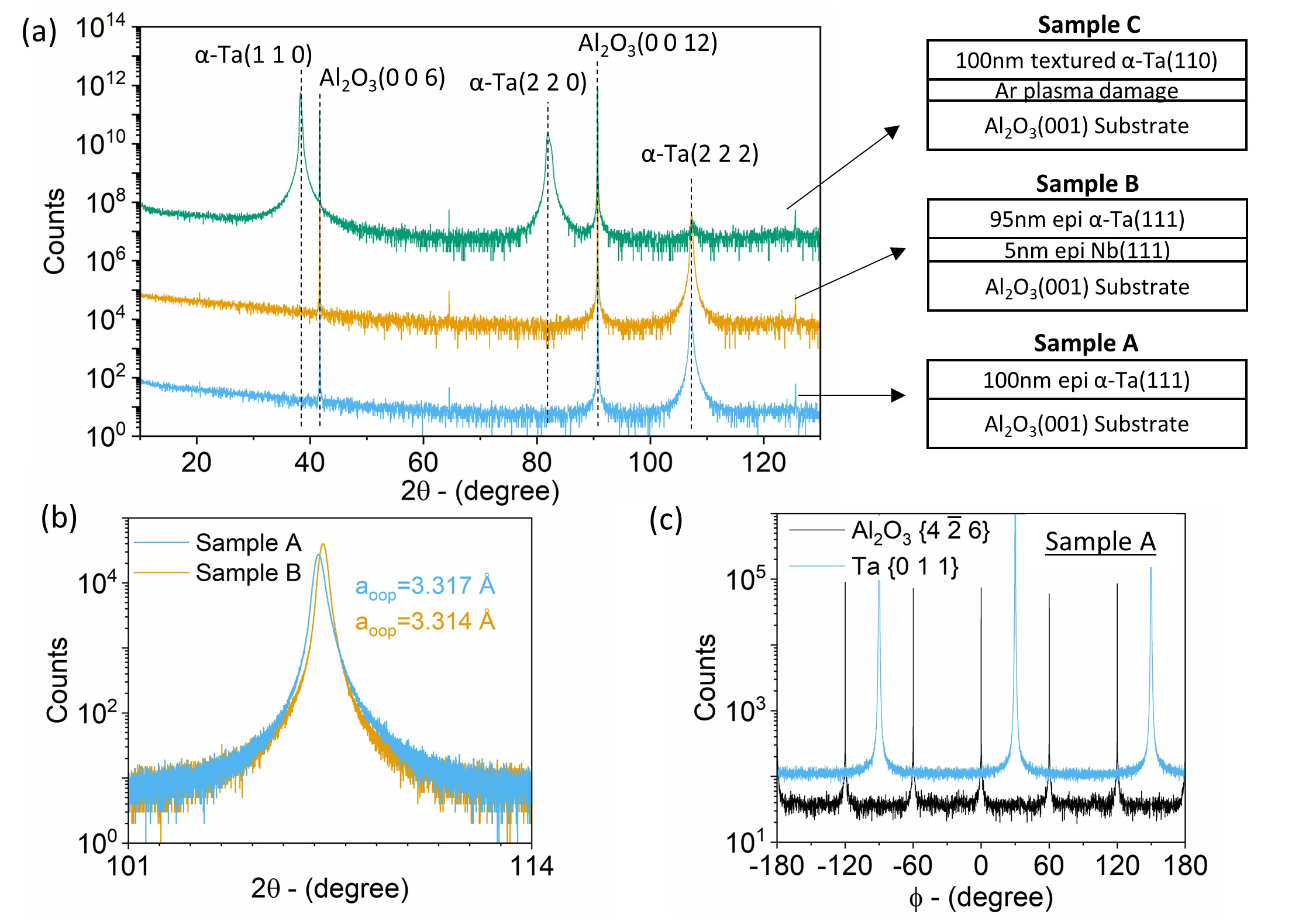} 
\caption{\label{XRD} (a) Coupled ($\omega-2\theta$), on-axis X-ray diffraction measurements of Samples A, B, and C. Samples A and B are epitaxial while sample C is textured (polycrystalline). Data are offset for clarity. (b) High-resolution scans of the $Ta(222)$ reflection from (a) showing comparison between Samples A and B and their corresponding out-of-plane lattice constants ($a_{oop}$). The difference in Ta lattice constant between Samples A and B is less than 1\% (c) $\phi$-scan (in-plane rotation) of off-axis reflections from the Al$_2$O$_3$(0001) substrate and $Ta(111)$ confirming the film is epitaxial and it's rotational alignment to the substrate. The indicated family of reflections from sapphire $Al_2O_3\{4 \bar2 6\}$ is in the cubic coordinate system.}
\end{figure*}

XRD measurements of Samples A, B, and C are included as Fig.\ref{XRD}. From the specular, coupled scan presented in Fig.\ref{XRD}(a), it is clear that Sample C grows primarily in the (110) orientation in contrast to Samples A and B, which grow in the (111) orientation. 

Figure \ref{XRD}(b) shows higher resolution scans of the $\alpha$-Ta(222) reflections from which the difference in the out-of-plane lattice constant of the Ta films grown with and without the Nb inter-layer is computed to be less than 1\%. Fig. \ref{XRD}(c) shows $\phi$- scans (in-plane sample rotation) of non-specular substrate and film peaks of Sample A, which confirm that the Ta is epitaxially arranged to the substrate having the same rotational relationship stated earlier for the epitaxial Nb. Non-specular XRD measurements of Sample C were also performed which showed in-plane texturing. Additional XRD data is included in Appendix C.

Figure \ref{AFM} includes atomic force micrographs taken from Samples A, B, and C. For Samples A and B, (the epitaxial Ta with and without the 5nm thick Nb interlayer), the surface morphology was smooth and shapeless with little difference between the two. For Sample C (the Ta film grown on plasma-treated substrate), elongated, grain-like features were observed, consistent with previous reports of textured Ta and Nb films \cite{Clavero2012,Jones2023}. The direction of the elongation reflects the in-plane texturing of the film where the same 6 rotational domains measured with XRD are observed. RMS roughness values extracted from AFM data is included in Table \ref{filmData_Ta} for these samples.

\begin{figure}[hbt!]
\includegraphics[width=0.48\textwidth]{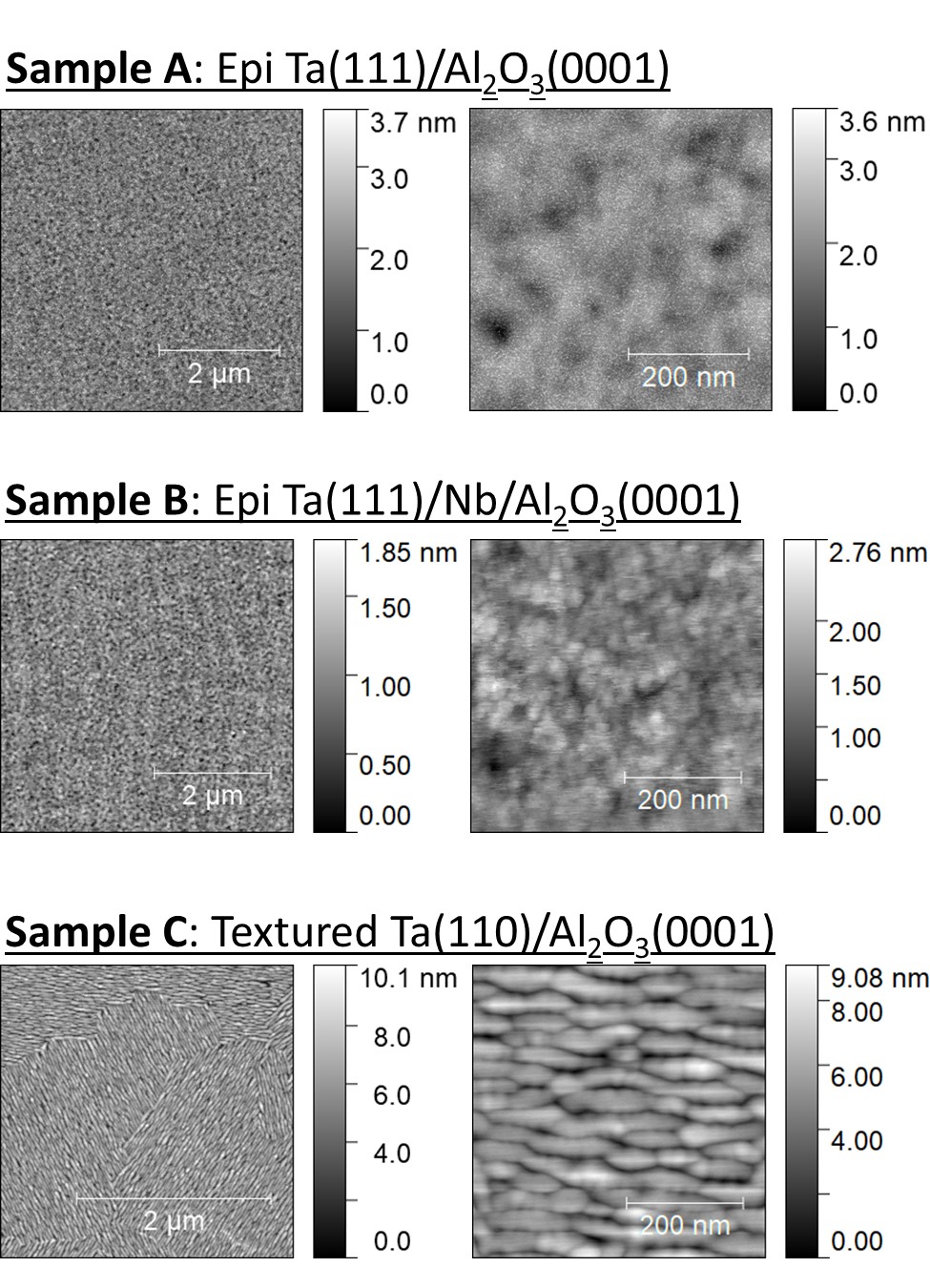} 
\caption{\label{AFM} Atomic force micrographs of samples A, B, and C showing the surface morphology of Ta films grown at the same temperature but nucleated with different interfaces.}
\end{figure}

\begin{figure}[hbt!]
\includegraphics[width=0.48\textwidth]{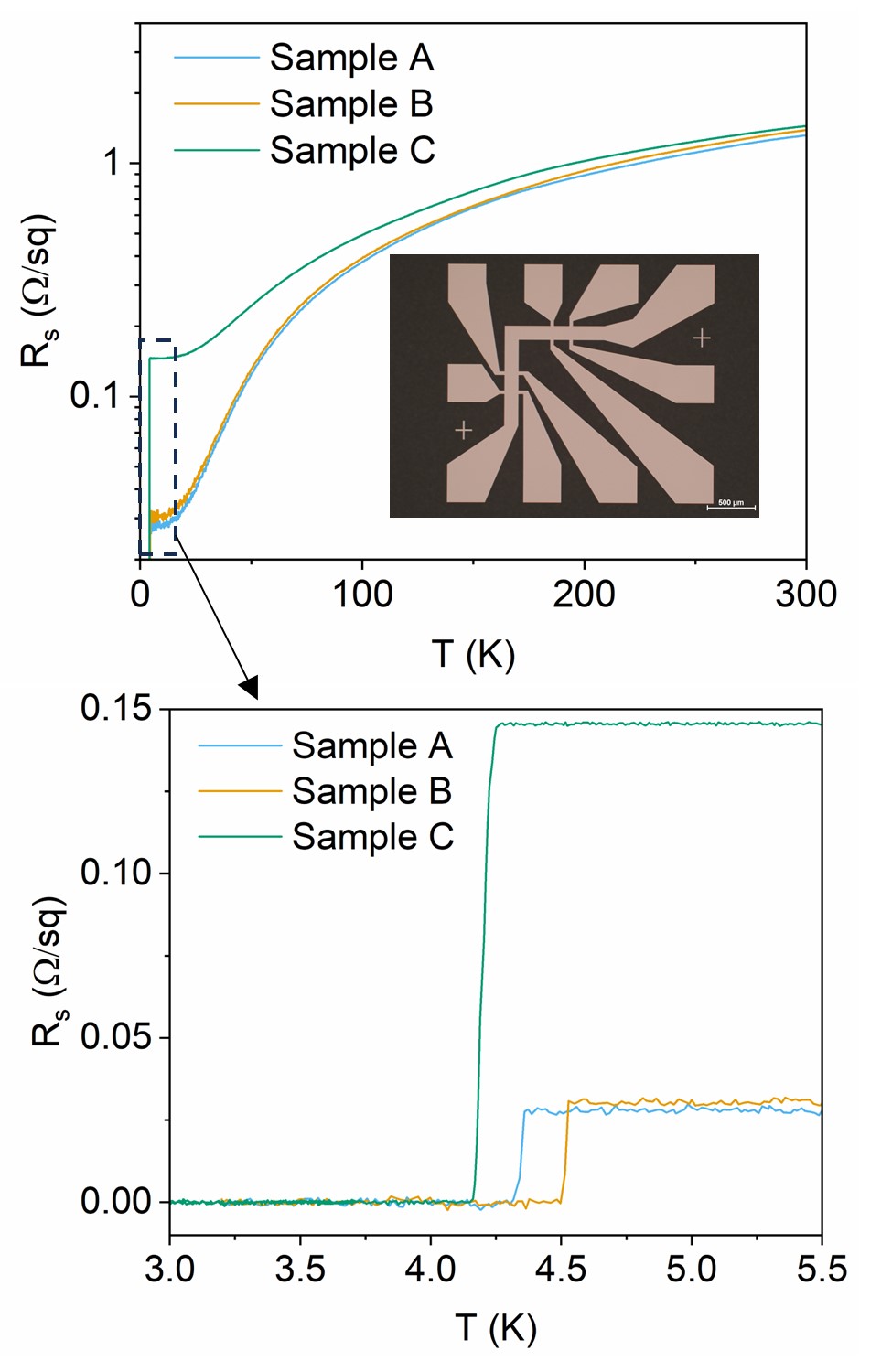} 
\caption{\label{DCTransport} Temperature-dependent resistance measurements of samples A, B, and C. The RRR of sample C is lower than A and B, likely as a result of its polycrystalline structure}
\end{figure}

Figure \ref{DCTransport} shows temperature dependent resistivity measurements of Samples A, B, and C. For the epi-$\alpha$-Ta(111) (Samples A and B), the 5nm Nb interlayer had little effect on film resistance, RRR, or $T_c$. The RRR of the textured Ta(110) film grown on the plasma damaged sapphire (Sample C) was lower, reflecting less structural order in the polycrystalline film. The $T_c$ of all $\alpha$-Ta films was between 4.2-4.5K and the complete data is included in Table \ref{filmData_Ta}.

\begin{figure*}[hbt!]
\includegraphics[width=1.0\textwidth]{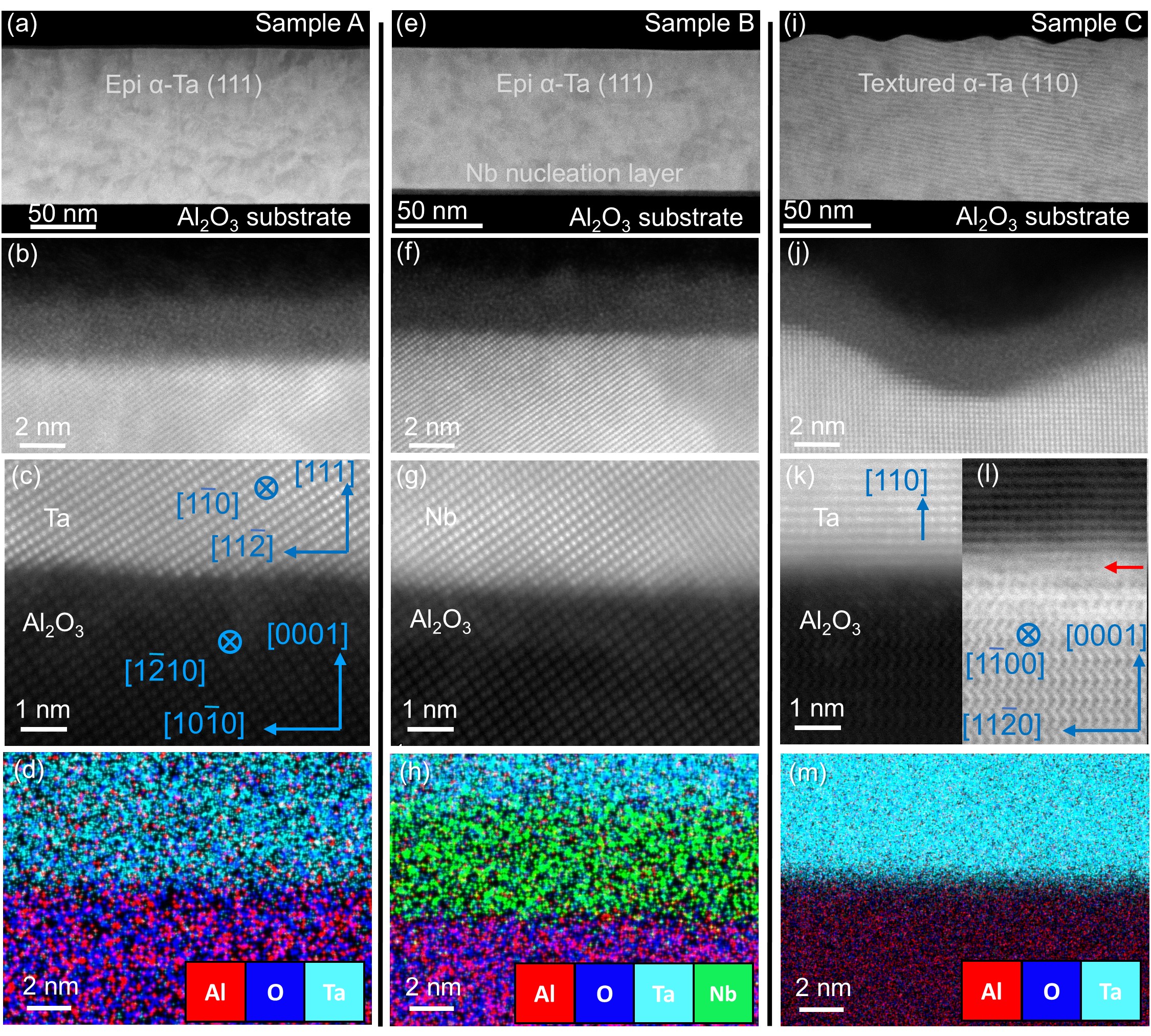} 
\caption{\label{TEM} Scanning transmission electron micrographs of Samples A(a-d), B(e-h), and C(i-m). (a, e, i) Low mag HAADF-STEM images. (b, f, j) Atomic resolution HAADF-STEM images of Ta surface. (c, g, k) Atomic resolution HAADF-STEM images of Ta (or Nb)/$Al_2O_3$ interface of Samples A, B, and C, (l) an atomic resolution ABF-STEM image of Sample C, and (d, h, m) corresponding EDS elemental distribution maps. The metal/sapphire interface is chemically abrupt in all three samples. Damage caused by Ar plasma treatment of the sapphire surface prior to Ta growth in Sample C is clearly visible in (l), appearing as a discontinuity of the lattice indicated by red arrow.}
\end{figure*}

The overall microstructure of Samples A, B, and C was further investigated by HAADF STEM imaging. Low-magnification STEM images (Figs. \ref{TEM} (a), (e), and (i)) show that both Samples A and B have flat surfaces, while Sample C exhibits obvious surface waviness consistent with AFM measurements. Nevertheless, the Ta surface oxide layers in Samples A, B, and C have similar thicknesses of approximately 3nm as shown in Figs. \ref{TEM}(b), (f), and (j). High-resolution STEM imaging shows a clear epitaxial relationship of $(Ta,Nb)[111]||Al_2 O_3 [0001]$ with $(Ta,Nb)[1 1 \bar2]||Al_2 O_3 [1 0 \bar1 0]$ in Samples A and B. The metal/substrate interfaces are abrupt and chemically sharp, as shown in EDS maps (Fig. \ref{TEM}(d), (h), and (m)). However, the Ar-plasma treatment of Sample C introduces damage to the sapphire surface, and the Ta film grows as a polycrystalline textured film with an approximately 1nm thick damaged $Al_2 O_3$ layer at the metal/substrate interface Fig. \ref{TEM}(l).

\section{Discussion}
Consider Samples A, B, and C as indicated in Table \ref{filmData_Ta}. All three films were grown on c-plane sapphire at the same temperature of 630$^\circ C$: Sample A is epi-$\alpha$-Ta(111) grown directly on sapphire, Sample B is epi-$\alpha$-Ta(111) grown on a 5nm thick Nb interlayer on sapphire, and Sample C is textured-$\alpha$-Ta(110) grown on plasma treated sapphire. The $Q_i$ of resonators made from Sample A are markedly lower than B and C. The microscopic mechanism of this loss is of interest, and some potential sources are discussed here in light of consistency with the results presented in this work. 

Firstly, the epi-$\alpha$-Ta(111) samples grown with and without the Nb nucleation layer (Samples A and B) are considered. Structurally, the Ta films are very similar as evidenced by the XRD and AFM data provided in Figs. \ref{XRD} and \ref{AFM} respectively. The measured resistivity, superconducting critical temperature, and RRR (see Table \ref{filmData_Ta}) for these two samples are also all very similar, yet the resonator $Q_i$ differs by about a factor of 60. Together, this suggests that the microwave loss in Sample A arises at the Ta/Sapphire interface and is not a result of the Ta(111) film structure, superconducting critical temperature, RRR, or surface morphology.

Next, consider the Ta film grown on the plasma treated substrate (Sample C). The resulting Ta film structure is much different from Samples A and B even though the growth temperature (630$^\circ C$) was the same. Here, the Ta film grew in the (110) orientation, the low energy surface of the BCC structure, as a result of plasma-induced damage to the substrate surface. The Ta film was confirmed to be polycrystalline, but with in-plane texturing that was observed with XRD, AFM, and TEM. 

Sample C showed a RRR of 9.9 which was reduced compared to the epitaxial Samples A and B (RRR > 40): an expected result from the more structurally defective film. AFM measurements show that the surface is also significantly rougher at 13.3\r{A} RMS roughness as compared to < 4\r{A} for Samples A and B. Structurally, Sample C is more defective and rougher than Samples A and B. Nonetheless, the $Q_i$ of resonators made from this film are > 40x improved compared to the Sample A which suggests that the loss arising from the interface is more significant than any effects having to do with differences in roughness and disorder in the Ta films.

Initially, the origin of the observed interface loss was thought to be either due to reactions between the Ta and the Al$_2$O$_3$ substrate or chemical contamination on the sapphire surface before growth, but neither explanation is fully consistent with the results. First addressing interface contamination: the substrates used for all samples presented in this study (with exception of Sample C) were prepared in the same way and thus had very similar levels of surface contamination before film growth. High $Q_i$ was observed for all Nb films grown, and also for the Ta films grown at lower temperature. Having surface contamination that only affects Ta grown at high temperature seems unlikely.

Regarding reactions between the Ta film and the sapphire substrate due to elevated growth temperature, no additional crystalline phases were observed in XRD measurements of Samples A, B, or C, and TEM results suggest that the metal/sapphire interface of all three samples is chemically abrupt to within a few atomic layers. Moreover, if metal/oxide reactions had occurred in Sample A and resulted in poorly performing resonators, one would expect the same to happen in Sample C, which was also Ta grown directly on sapphire at the same temperature. 

These results emphasize the importance of substrate surface preparation before metal growth to obtain devices with low microwave loss, but not necessarily just to reduce chemical contamination at the metal/substrate interface. Plasma treatments employing oxygen and argon to 'clean' the substrate surface are very commonly used for this application \cite{Place2021,Jones2023}, but these treatments also induce damage to the substrate surface before growth, which affects both the metal/substrate interface and the resulting structure of the metal film. The results presented here suggest that, at least in the case of Ta on c-plane sapphire, the structural change to the interface plays a more significant role in reducing MW loss than any effects from reducing chemical contamination.

Before speculating on potential microscopic origins of the interface-dependent loss, it is emphasized that the lossy resonators made from epi-$\alpha$-Ta(111) grown directly on c-plane sapphire display low $Q_i$ which is essentially power independent in the range studied (1,000,000 to <1 photon occupation in the resonator). This observation suggests that the source of microwave loss is not saturable TLS, but rather a power-independent mechanism such as quasiparticles\cite{Aumendato2023}, piezoelectricity\cite{Scigliuzzo2020,McRae2021}, vortex motion\cite{McRae2020b}, or magnetism. 

For the case of clean Ta films at dilution refrigerator temperatures (35mK in this work), thermally activated quasiparticles are exponentially suppressed \cite{Aumendato2023} and are not expected to be an observable source of MW loss in CPW resonators. Suppression of superconductivity at the Ta/sapphire interface could in principle result in an equilibrium quasiparticle density responsible for the observed loss, but the mechanism of suppression is not clear. It has long been established that the gap energy of a BCS superconductor in contact with a trivial insulator is only negligibly reduced\cite{DeGennes1964}. Even normal metals such as gold (Au) tend to proximitize and high performing microwave devices have been demonstrated using Au capped Nb and Ta\cite{Bal2024,Chen2024}. For these reasons, trivial suppression of superconductivty at the interface to the point that the observed MW loss may be explained by equilibrium quasiparticles seems unlikely.

Song et al.\cite{Song2009} studied the microwave response of vorticies to applied MW fields in different superconducting films, concluding that differences in vortex dynamics between the films resulted in different contributions to overall MW loss in CPW resonators. It is conceivable that the vortex dynamics are different in Samples A, B, and C in this work owing to the differences in film and interface structure. The critical magnetic field of bulk Ta ($B_{c0}$) is above 750 Gauss\cite{Hinrichs1961}, which is much greater than what could be expected for this measurement condition. However, when thin strips of superconductor are cooled through $T_c$ in applied perpendicular field, vorticies appear at significantly lower threshold magnetic flux density which was shown by Stan et al.\cite{Stan2004} to be essentially material independent and well approximated by $B_m \approx \Phi_0/W^2$, where W is the width of the strip and $\Phi_0$ is the magnetic flux quantum. 

Given the CPW centerline dimension of $6\mu m$ used in this work, a corresponding threshold $B_m=0.57 G$ is obtained; a value comparable to the earth's magnetic field ($\approx0.5 G$). Thus, it is conceivable that trapped vorticies exist in the CPW resonators studied here; however, the residual field from the earth will likely not be aligned perpendicular to the sample, magnetic shielding was used for all measurements here, and the lossy resonators were measured to be essentially the same in multiple cooldowns using multiple chips as well as different setups. In the report by Stan et al., vorticies were not observed without an applied field during cooldown, even in 100nm wide superconducting strips. Moreover, even if the observed difference in MW loss were due to differences in vortex dynamics between the films, the microscopic mechanism remains unclear.

Stability of the tantalum surface oxide is often cited as the reason for performance improvements in MW devices employing Ta as compared to Nb. For both Ta and Nb oxides, the most stable is the 5+ oxidation state (Nb,Ta)$_2$O$_5$ , though sub-oxides are possible, tend to be more abundant in Nb-oxide as compared to Ta-oxide\cite{Place2021,Crowley2023}, and are often cited as a dominant source of MW loss in Nb based devices. Detailed growth studies of Nb and Ta on A-plane sapphire suggest that the initial structure of the deposited metal is likely determined by bonding of Nb or Ta to oxygen on the sapphire surface\cite{Oderno1998,Ward2003,Grier2000}, thus Nb and Ta bonding with oxygen will also occur at the interface with sapphire, though is limited in extent to just several atomic layers. 

It is conceivable that the electronic structure resulting from this interfacial bonding could be a source of MW loss arising from electronic states or potentially magnetism. Some sub-oxides of Nb, W, and V are known to display magnetic behavior\cite{Cava1991,Zheng2020,Cezar2014,Parida2010}. Certain nanostructures of Ta$_2$O$_5$ have also been shown to be ferromagnetic\cite{Lu2024}, exemplifying the sensitivity of electrical and magnetic properties to physical structure. 

The observed differences in microwave loss with interface structure may be caused by differences in oxygen bonding across the Ta/Sapphire interface. For future work, \textit{In-situ} RHEED and X-ray photoelectron spectroscopy (XPS) studies may be used to study the chemical bonding between tantalum and oxygen in the first few monolayers of growth as the interface is formed\cite{Fujii1996,Beran2019}. Lattice-site specific bonding of tantalum to oxygen may be studied using X-ray standing wave (XSW) techniques\cite{Chen2024b} which may prove insightful comparing the ordered, epitaxial interface of Sample A to the disordered interface of Sample C, for example. Notably, all films discussed in this work were grown on a sapphire surface that underwent a high temperature anneal in vacuum in order to obtain the same surface chemistry and Al-O stoichiometry. The oxygen content of this surface may be experimentally controlled by ion bombardment, oxygen dosing, or thermal annealing for example\cite{Gillet1992} which may correlate to microwave loss.
 
Stageberg et al. studied Josephson junctions made using the ferromagnetic insulator Ho(OH)$_3$ as a tunneling barrier and observed threefold splitting in the tunneling conductance peaks\cite{Stageberg1985}. These spectra were qualitatively explained by DeWeert and Arnold\cite{DeWeert1985,DeWeert1989} where it was shown that superconducting-quasiparticle interface states arise at energies within the superconducting gap due to the interface with an insulating ferromagnet. A similar situation (i.e. quasiparticle-like states within the superconducting gap) arises considering a superconductor containing magnetic impurities\cite{Balatsky1997,Fominov2016}. The epitaxial interface between Ta and c-plane sapphire may be nontrivial, and quasiparticle interface states could certainly give rise to MW loss depending on their energy. If the interface is ferromagnetic, MW loss could also occur through magnetic damping\cite{Gilbert2004,Hickey2009}.

Bulk sapphire is not piezoelectric, though breaking of inversion symmetry at the Al$_2$O$_3$(0001) surface has been shown in theory to lead to piezoelectricity \cite{Georgescu2019}. Park et al. \cite{Park2022} induce piezoelectricity in non-piezoelectric CeO$_2$ through the rearrangement of oxygen vacancies by introducing Gd impurities. Thus, piezoelectricity could potentially exist at the metal/sapphire interfaces of the samples used in this work. Differences in the preparation conditions between Samples A, B, and C may result in different interfacial strain, which could affect the piezoelectric properties of the interface leading to the observed differences in MW loss.

\section{Conclusion}

Structural and electrical properties of Nb and Ta thin films deposited on c-plane sapphire were systematically studied for superconducting microwave device applications as a function of growth temperature. MW loss was quantified using CPW resonators, and while resonators made from Nb films grown in the temperature range 20-730$^\circ C$ were found to be high performing, only Ta films grown at lower temperature (including a sample grown at room temperature composed of mostly $\beta$-Ta) demonstrated high quality factors. Epitaxial Ta(111) films grown directly on Al$_2$O$_3$(0001) substrates were found to display poor MW performance, which contrasts with structural characterization as well as $T_c$, resistivity, and RRR measurements which suggest the films are high quality.

To further explore the loss in the poorly performing devices, the Ta interface with sapphire was tested by either growing a 5nm thick epitaxial Nb interlayer, or by plasma-treating the sapphire surface before Ta growth. Improvements to resonator quality factors over 10x were observed for both approaches, suggesting that the loss in the epi-$\alpha$-Ta(111) resonators arises at the Ta/sapphire interface, and that loss may be mitigated with surface treatments. All these data together suggest that the observed loss in epi-$\alpha$-Ta(111) grown directly on sapphire is not simply due to chemical contamination or additional reacted phases, but rather the structure of the interface. Some potential microscopic mechanisms of MW loss were discussed which include quasiparticle-interface states due to nontrivial electronic or magnetic interface structure, strain-sensitive interface piezoelectricity, and vortex dynamics.

\begin{acknowledgments}
We wish to acknowledge Dominic Goronzy, Carlos G. Torres-Castanedo, Corey Rae McRae, Nikolay Zhelev, and David Garcia for useful discussions and important feedback, and Mustafa Bal for fabrication process development. We also thank Chris Yung and Peter Hopkins for a critical reading of this manuscript.

This material is based upon work supported by the U.S. Department of Energy, Office of Science, National Quantum Information Science Research Centers, Superconducting Quantum Materials and Systems Center (SQMS) under contract no. DE-AC02-07CH11359.

\end{acknowledgments}

\section*{Competing Interests and Data Availability Statement}

All authors declare no financial or non-financial competing interests. The datasets used and/or analyzed during the current study available from the corresponding author on reasonable request. This work is a contribution of the U.S. government and is not subject to U.S. copyright. Certain commercial equipment, instruments, or materials are identified in this paper to foster understanding. Such identification does not imply recommendation or endorsement by the National Institute of Standards and Technology, nor does it imply that the materials or equipment identified are necessarily the best available for the purpose.

\section*{Appendix A: Additional Epitaxial T\lowercase{a} Films}
The most important finding in this work, namely that the highest quality epitaxial Ta films yield lossy resonators when grown directly on sapphire but are high performing when grown on a thin, epitaxial Nb inter-layer has been reproduced in samples not included in the main text. Table \ref{additionalTable} includes transport data for 4 additional epitaxial $\alpha$-Ta samples along with those already included in the main text. The two additional Ta films grown directly on sapphire yielded resonators with $Q_i < 50k$, while the films grown on an epitaxial Nb interlayer yield resonators with $Q_i > 500k$, consistent with the conclusions outlined in the main text.

The crystalline structure of these epitaxial $\alpha$-Ta films is very similar, and all test resonators underwent the same BOE cleaning procedure before measurement, however there are some important differences that should be emphasized. Sapphire wafers having both double- and single-side polishing (DSP/SSP) were used as well as several different vendors. In addition to any inherent differences between the substrates, the backside polish affects the wafer mounting configuration in the growth chamber and thus the surface growth temperature as well.

Samples having two additional surface preparations before film growth are also included; one with a chemical clean in stabilized Piranha solution held at $80^\circ C$ for 10 minutes, and another annealed in air at $1100^\circ C$ in order to obtain a terraced sapphire surface\cite{Dwikusuma2002,Vlasov2016} which was confirmed using AFM. All samples were annealed and outgassed at $880^\circ C$ before cooling to the indicated growth temperature for film deposition.

\section*{Appendix B: RHEED characterization of the damaged sapphire surface}
\begin{figure*}
\includegraphics[width=1.0\textwidth]{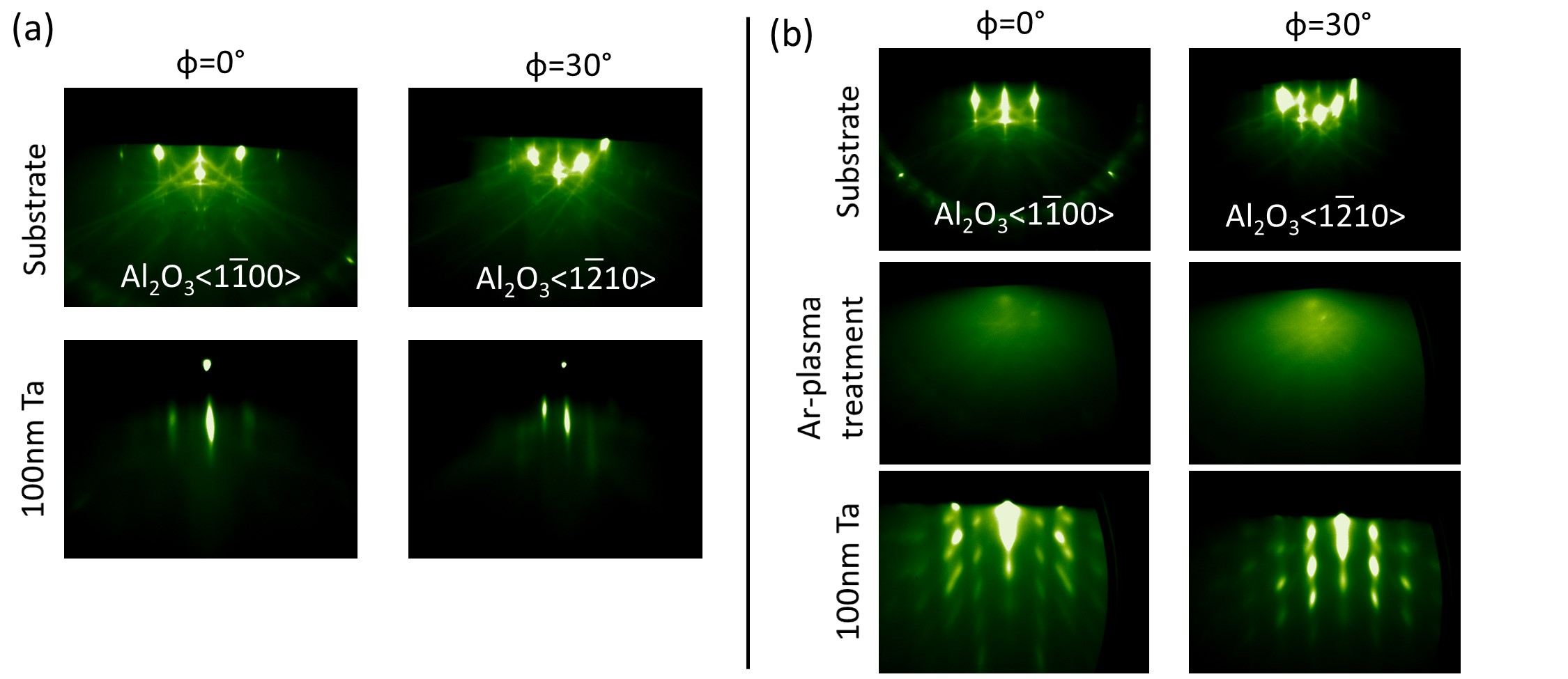}
\caption{\label{RHEED} (a) RHEED patterns of the starting $Al_2O_3(0001)$ substrate surface and resulting 100nm Ta film grown at $630^\circ C$ (Sample A). The sapphire surface is crystalline and the Ta film is smooth and epitaxial. (b) RHEED images showing the effect of plasma damage to the sapphire surface and the resulting textured Ta film also grown on this surface at $630^\circ C$ (Sample C). After Ar plasma treatment, the bright diffraction pattern from the sapphire substrate changes to a diffuse background indicating damage to the ordered surface. The Ta film grown on this damaged surface is textured both out-of-plane and in-plane as measured by XRD. RHEED patterns were taken \textit{in-situ} without exposing the samples to air.}
\end{figure*}

Ar plasma exposure is expected to both etch the sapphire substrate and induce structural damage to the sapphire surface. The structural damage to the starting surface of Sample C was observed using \textit{in-situ} RHEED, where the bright diffraction pattern of the sapphire observed initially became dim (though still observable) after plasma exposure, and was accompanied by an increase in the diffuse background, indicating a less ordered surface. This diffuse RHEED pattern did not improve following high vacuum annealing at $880^\circ C$ for 2 hours and dropping to the growth temperature of $630^\circ C$. RHEED patterns of this sample after Ta growth were consistent with a textured film having both out-of-plane and in-plane texture. This texture was confirmed with XRD after the sample was cooled and removed from the growth chamber. 
Figure \ref{RHEED} shows \textit{in-situ} RHEED images of the starting sapphire surfaces of Samples A and C, the damaged sapphire surface from Sample C, and the resulting 100nm thick Ta films.

\section*{Appendix C: XRD pole figures}
\begin{figure*}
\includegraphics[width=1.0\textwidth]{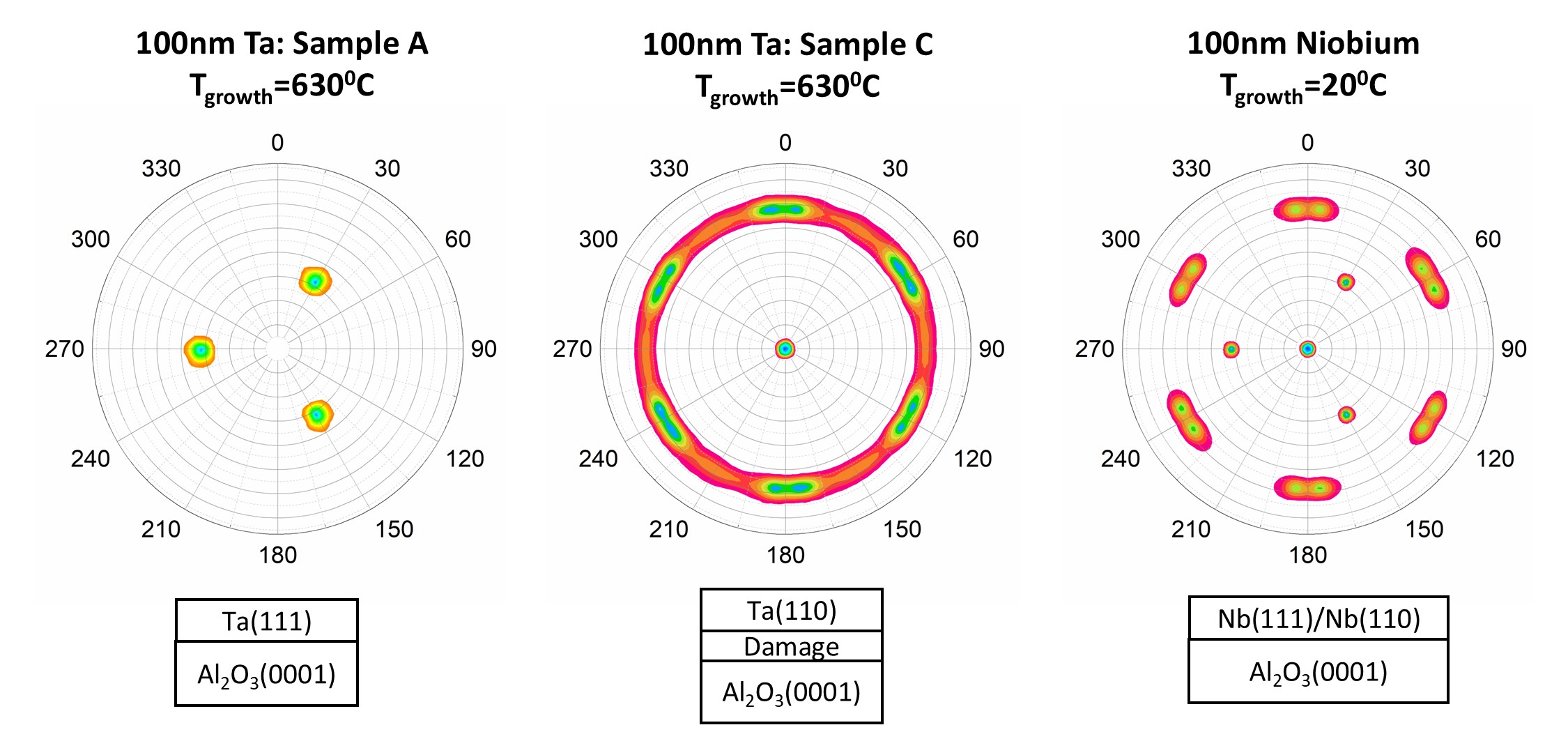}
\caption{\label{Polefigs} X-ray diffraction pole figures of the Ta/Nb (110) reflections for epitaxial (Sample A), textured (Sample C) 100nm thick Ta films grown at $630^\circ C$ on $Al_2O_3(0001)$, as well as the Nb film grown at room temperature. The three peaks observed for Sample A confirm the film has a single crystalline orientation and is epitaxial to the sapphire substrate.}
\end{figure*}

In addition to the non-specular measurements included in Fig. \ref{XRD}(c), pole figure measurements were also performed using the same diffractometer configuration outlined in the Methods section of the main text. These data complement the specular ($\omega-2\theta$) measurements included in Figs. \ref{NbTa_XRD} and \ref{XRD} and are used here to examine epitaxy and in-plane texturing. Figure \ref{Polefigs} shows polar plots of the Ta/Nb(110) reflections (2$\theta = 38.3^\circ$) for three samples included in the main text: the epitaxial Ta film of Sample A, the polycrystalline textured film of Sample C, and the Nb film grown at room temperature.

Turning first to Sample A, three distinct peaks are observed separated by a polar angle of $120^\circ$ corresponding to the Ta(110), (101), and (011) crystal planes of the (111) oriented Ta film. This measurement confirms that the film is epitaxial with a single crystalline orientation. Turning next to Sample C, a clear peak is observed at zero azimuthal angle, consistent with specular measurements which had shown the film is (110) oriented. The off-specular peaks from this film are observed to have some structure, with peak maxima slightly misaligned from the sapphire high symmetry directions which confirms this film is highly textured out of plane in the (110) orientation with some in-plane texturing as well. The Nb film grown at room temperature has similarity to both Samples A and C, containing both epitaxial (111) grains and (110) grains having similar in-plane texture to Sample C.

\section*{Appendix D: AFM measurements of growth temperature series}
\begin{figure*}
\includegraphics[width=1.0\textwidth]{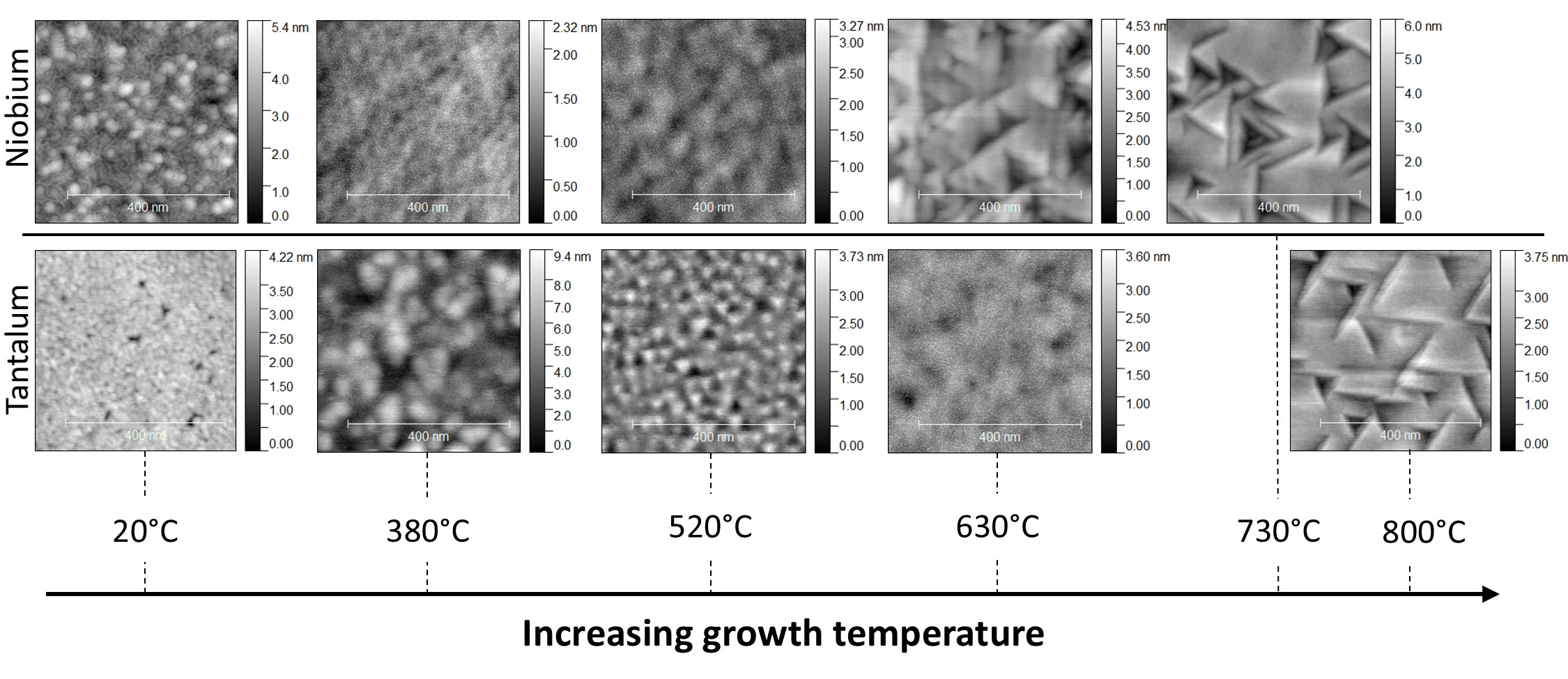}
\caption{\label{Supp_AFM} Atomic force micrographs of 100nm thick Nb and Ta films grown on $Al_2O_3(0001)$ at varying substrate temperatures. All films were grown on clean, crystalline $Al_2O_3(0001)$ surfaces without any plasma treatment.}
\end{figure*}

Atomic force microscopy was performed on all films presented in this study. Figure \ref{Supp_AFM} shows AFM images of the Nb and Ta films grown at varying substrate temperature. At higher growth temperatures, triangular features are observed in the surface morphology. For Nb, the triangular morphology was observed for both the samples grown at $520^\circ C$ and $630^\circ C$. For Ta, triangular morphology is also observed, though the required temperature is higher than that for Nb.

\section*{Appendix E: Quality factor measurement details and power dependence}
The procedure used for measuring internal quality factor and estimating photon occupancy of the hanger resonators is as follows: first, the individual resonators are mapped out manually using a vector network analyzer (VNA) to obtain the approximate resonance frequency and needed frequency span to encompass the resonance line-shape. Transmission (S$_{21}$) spectra are then acquired over a range of VNA output powers. These data are then fit to the diameter correction method model\cite{Khalil2012}, which for hanger resonators reduces to\cite{McRae2020b}:

\begin{equation}
S_{21}(f)=1-\frac{Q/\hat{Q}}{1+2iQ\frac{f-f_0}{f_0}}
\end{equation}

In this expression, $\hat{Q}=Q_ce^{-i\phi}$, $Q$ is the total quality factor defined by $Q^{-1}=Q_i^{-1}+Re\{\hat{Q}^{-1}\}$,  $f_o$ is the resonance frequency, $Q_i$ and $Q_c$ are the internal and external quality factors respectively, and $\phi$ is the correction angle representing asymmetry in the Lorenzian lineshape, More details on the derivation of these expressions and physical origin of the correction angle $\phi$ is provided by \cite{Khalil2012,Rieger2023}.

The power used to probe the resonators may be expressed as an average photon number $n$ in the cavity. For a hanger resonator driven on resonance, this photon number is given by the following expression \cite{Bruno2015}.
\begin{equation}
n=\left(\frac{Q^2}{\pi Q_c h f_0^2}\right)P 
\end{equation}
where $h$ is Planck's constant and $P$ is the incident power at the device.  The parameters $f_0$, $Q$, and $Q_c$ are extracted from the fit to the DCM model above. $P$ is estimated using the power output of the VNA having a dynamic range of +20~dBm to -87~dBm, with the known attention of $\approx$80~dB from the VNA to the mixing chamber of the dilution refrigerator.  This includes 70~dB of attenuation from cryogenic attenuators and an estimated 10~dB of cabling losses.



\begin{figure*}
\includegraphics[width=1.0\textwidth]{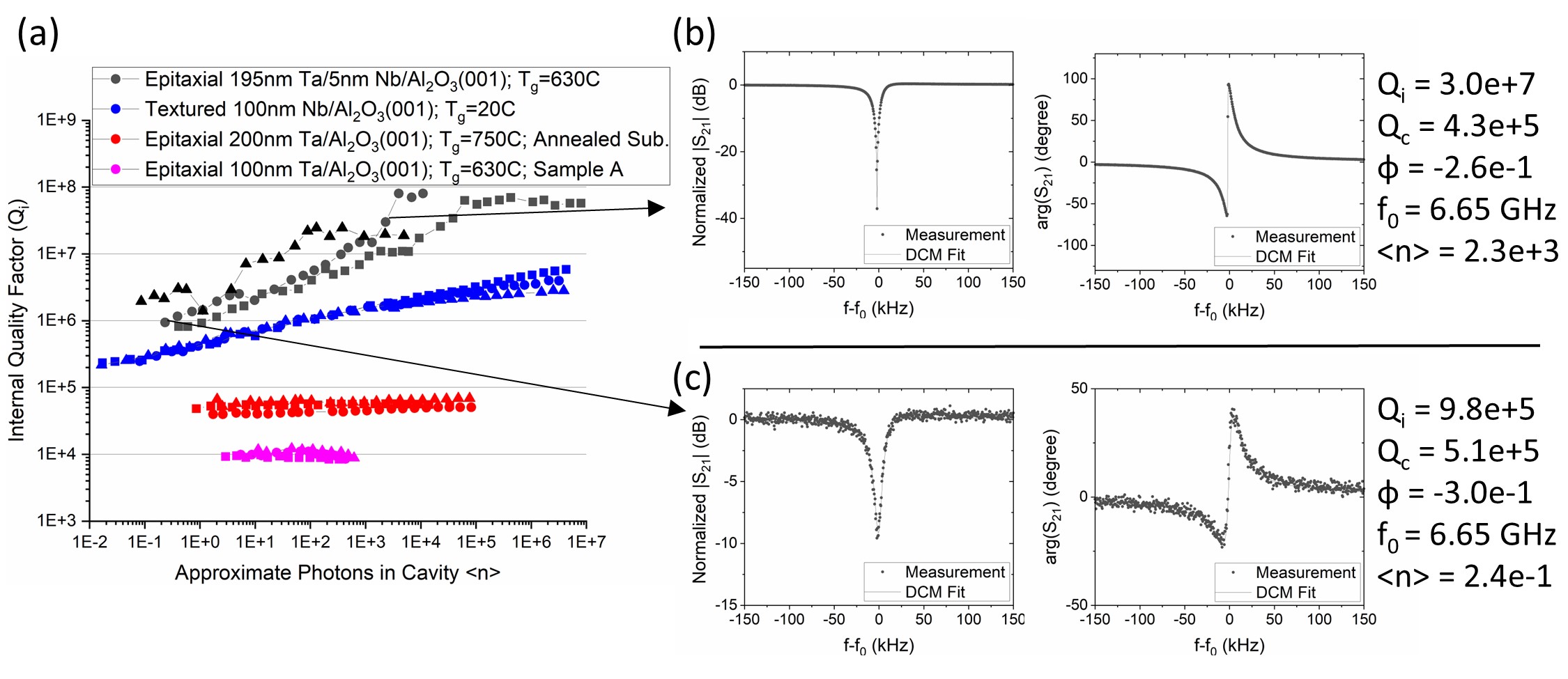}
\caption{\label{Supp_Qi} (a) Internal quality factor plotted against estimated photon number for several resonator chips presented in this work. The different symbols (triangle, square, and circle) represent distinct resonators on the same chip. $Q_i$ of the worst performing resonators (epitaxial Ta grown directly on Al$_2$O$_3$(0001)) is relatively flat with power. Example $S_{21}$ data along with fits to the DCM model for some of the best performing resonators is shown at (b) high, and (c), low power.}
\end{figure*}

The power dependence of resonator $Q_i$ is commonly used to extract the contribution of saturable loss sources or two-level systems (TLS) which have been suggested to be a source of decoherence in state-of-the-art superconducting qubits\cite{McRae2020b}. Figure \ref{Supp_Qi} shows resonator $Q_i$ vs estimated photon number for several chips that were presented in this work as well as some example $S_{21}$ traces and fits to the DCM model. The data included in Fig.\ref{Supp_Qi}(a) was simply selected in order to provide representative curves while avoiding overlapping data sets. Throughout this work, a total of 16 chips were measured in this way, though data for only 4 chips is included in this plot for clarity. For the best performing resonators (epitaxial Ta grown on a Nb nucleation layer), $Q_i$ varies by nearly two orders of magnitude in the range studied demonstrating the contribution of TLS to the low power loss. In contrast, $Q_i$ of the worst performing resonators (epitaxial Ta grown directly on Al$_2$O$_3$(0001)) show a weak or negligible power dependence in the range accessible.

\begin{table*}
\begin{ruledtabular}
\begin{tabular}{r|cccccc}

$Al_2 O_3 (0001)$ Substrate/Polish. & Surface Prep & Film & Growth Temp. & RRR & $T_c$ ($K$) & Resonator mean $Q_i$\\
  &   &   &  ($^\circ C$) &   &   & (Low-power) \\\hline

Vendor A, SSP (Sample A)\footnote[1]{Included in main text} & Solvent Clean/HV Anneal & 100nm Ta/$Al_2O_3$ & 630 & 47 & 4.35 & 8.3k\\
Vendor A, SSP\footnotemark[1] & Solvent Clean/HV Anneal & 100nm Ta/$Al_2O_3$ & 520 & 29 & 4.24 & 7.4k\\
Vendor B, SSP & Stabilized Piranha/HV Anneal & 100nm Ta/$Al_2O_3$ & 800 & - & - &  < 20k \\
Vendor C, DSP\footnote[2]{Wafer underwent additional Josephson junction processing steps} & 3 hr.@1100C in Air/HV Anneal & 200nm Ta/$Al_2O_3$ & 750 & - & - & 32k\\
Vendor A, SSP (Sample B)\footnotemark[1] & Solvent Clean/HV Anneal & 95nm Ta/5nm Nb/$Al_2O_3$ & 630 & 45 & 4.35 & 520k\\
Vendor A, DSP & Solvent Clean/HV Anneal & 195nm Ta/5nm Nb/$Al_2O_3$ & 645 & 78 & 4.33 & 1.3M\\
Vendor A, DSP\footnotemark[2] & Solvent Clean/HV Anneal & 195nm Ta/5nm Nb/$Al_2O_3$ & 645 & 52 & 4.29 & 740k\\

\end{tabular}
\end{ruledtabular}
\caption{\label{additionalTable}Measured properties of epitaxial $\alpha$-Ta films grown with and without a Nb nucleation layer on c-plane sapphire}
\end{table*}

\clearpage
\bibliography{main}

\end{document}